# p-type doping in CVD grown $MoS_2$ using Nb

M. Laskar[1], D. N. Nath[1a], L. Ma[2], E. Lee[1], C. H. Lee[1], T. Kent[3], Z. Yang[3], Rohan Mishra[4], Manuel A Roldan[4], Juan-Carlos Idrobo[4], Sokrates T. Pantelides[4], Stephen J. Pennycook[4], R. Myers[3], Y. Wu[2] and S. Rajan[1b]

[1]Department of Electrical and Computer Engineering
[2]Department of Chemistry
[3]Department of Material Science and Engineering
The Ohio State University, Columbus, OH, 43210

[4]Materials Science and Technology Division
Oak Ridge National Laboratory

**Abstract:**

We report on the first demonstration of p-type doping in large area few-layer films of (0001)-oriented chemical vapor deposited (CVD) $MoS_2$. Niobium was found to act as an efficient acceptor up to relatively high density in $MoS_2$ films. For a hole density of 4 x $10^{20}$ $cm^{-3}$ Hall mobility of 8.5 $cm^2V^{-1}s^{-1}$ was determined, which matches well with the theoretically expected values. XRD and Raman characterization indicate that the film had good out-of-plane crystalline quality. Absorption measurements showed that the doped sample had similar characteristics to high-quality undoped samples, with a clear absorption edge at 1.8 eV. This demonstration of p-doping in large area epitaxial $MoS_2$ could help in realizing a wide variety of electrical and opto-electronic devices based on layered metal dichalcogenides.

[a] M. Laskar, D. Nath, L. Ma and E. Lee contributed equally to this work.
[b] Corresponding author. Email: rajan.21@osu.edu

Transition metal dichalcogenides (TMDs) such as $MoS_2$, $WS_2$, $WSe_2$ etc. have recently attracted widespread attention for a variety of next-generation electrical[1,2] and optoelectronic[3] device applications including low cost, flexible and transparent electronics[4,5]. These layered materials provide ultra-high confinement and are intrinsically 2-dimensional in nature, which is therefore promising for highly scaled vertical transistor topologies[6]. Besides, through van der Waals epitaxy, they circumvent limitations such as lattice mismatch in heterostructure growths of conventional semiconductors. Devices including field effect transistors (FET) with excellent on/off ratio and high current densities have been reported[1,7] using flakes of $MoS_2$ mechanically exfoliated from bulk geological samples. More recently, large area (0001) oriented $MoS_2$ with excellent crystalline and structural qualities grown by CVD on (0001) sapphire was reported[8]. Such CVD-grown $MoS_2$ eliminates the limitations associated with the commonly used exfoliated approach such as control on thickness and area, and are therefore viable for large scale device fabrication.

The use of Niobium (Nb) as substitutional impurity on the metal site to get p-type conductivity was reported several decades ago[9,10,11] for materials such as $MoS_2$ and $WSe_2$. However, material properties were not described, and the measurements were done for large (bulk) crystals. In contrast, few reports exist on p-doping in thin film (or even exfoliated) $MoS_2$. Using back-gating and liquid-gating approaches, p-type conductivity had been electrostatically achieved[12,13] on $MoS_2$ mechanically exfoliated from geological samples. However, the absence of p-doping using an acceptor dopant in epitaxial (and even mechanically exfoliated) $MoS_2$ has prevented demonstration of $MoS_2$-based bipolar devices such as heterojunction bipolar transistors (HBT), LEDs, photodetectors, etc. In this work, we show that as predicted by density functional theory based calculations[14], Nb is indeed an efficiency acceptor in $MoS_2$,

We followed a vapor deposition method for the growth of $MoS_2$[15] that was demonstrated to lead to high crystalline quality $MoS_2$ in previous work [8]. A series of three (0001) sapphire samples were prepared for $MoS_2$ growth by e-beam evaporation of Mo (2.5 nm)/Nb/Mo (2.5 nm) (schematic in Fig. 1) layers with thickness of Nb varied as 0.3 nm (sample A), 0.2 nm (Sample B) and 0 nm (control Sample C, undoped). The typical dimensions of the samples were 1-2 cm x 0.7 cm, with the breadth of the samples being limited by the size of the quartz tube used for the deposition. The sample was then subjected to sulfurization in CVD chamber at $900^0$C for 10 minutes. Further details of the CVD growth of undoped $MoS_2$ on sapphire were reported in [8]. The conditions optimum for growing large area undoped $MoS_2$ with excellent structural and surface qualities as reported in [8] were used in this study to grow all the samples. The thickness of $MoS_2$ grown under these conditions was found to be 10 nm from high resolution transmission electron micrograph (TEM)[8].

A Ni (30 nm)/Au (50 nm)/Ni (30 nm) metal stack was deposited by e-beam evaporation to form Ohmic contacts to Nb-doped $MoS_2$ (samples A and B). Devices were isolated by etching $MoS_2$ using $BCl_3$/Ar-based inductively coupled plasma/reactive ion (ICP-RIE) etch chemistry. Hall measurements were performed using standard van der Pauw pads, and both samples A and B had positive Hall coefficients indicating hole transport. Fig. 2 shows the temperature dependent hole concentration and hole mobility for sample A. No carrier freeze-out was observed even at 20 K indicating that degenerate p-type doping had been achieved. The room temperature hole mobility measured for sample A was 0.5 $cm^2$/Vs with hole density of $N_A=2x10^{21}$ $cm^{-3}$. Negligible dependence of hole mobility on temperature was observed. In comparison, sample B with reduced Nb thickness (0.2 nm) was found to exhibit a room temperature hole mobility of 8.5 $cm^2$/Vs with a corresponding p-type charge density of $4x10^{20}$ $cm^{-3}$. From Transfer Length Method (TLM) measurements, a low contact resistance of 0.6 Ωmm was extracted for sample B. The sheet resistance extracted from TLM was 1.8 kΩ/□ which was found to match that obtained from Hall measurement. The significant improvement in hole mobility (8.5 $cm^2$/Vs from 0.5 $cm^2$/Vs) with a reduction in p-doping density indicates that the mobility is limited mainly by ionized impurity scattering at such high degenerate doping densities.

A simple estimate was made for impurity scattering limited hole mobility in bulk $MoS_2$ using 3D formalism, given by $\mu_h = q\,\tau_{eff}/m^*$, where $\mu_h$ is hole mobility in $MoS_2$ while $\tau_{eff}$ and m* The momentum scattering time $\tau_m(N_A,E_F)$ at Fermi level $E_F$ was calculated using[16]:

$$\frac{1}{\tau_m(N_A,E_F)} = \frac{\pi\, N_A}{\hbar}\left(\frac{q^2 L_{TF}{}^2}{\varepsilon_0 \varepsilon_s}\right)^2 g_c(E_F) \qquad (1)$$

Here, k is Boltzmann's constant, ħ is reduced Planck constant, $\varepsilon_s$ is the dielectric constant[6] of $MoS_2$ (=3.3) and T is temperature. $g_c(E)$ in equations (1) is the 3-dimensional density of states for holes and $L_{TF}$ is Thomas-Fermi screening length for degenerate gas given by, $L_{TF} = \sqrt{\frac{\varepsilon_s}{q^2 g_c(E_F)}}$

Fig. 4 shows the hole mobility as a function of acceptor density ($N_A$) in p-$MoS_2$, and shows fairly good agreement with experiment, considering the large uncertainty in various parameters used in the model.

The material and structural quality of the $MoS_2$ samples was assessed using Raman spectroscopy, high resolution XRD scans and optical absorbance measurements. The undoped $MoS_2$ (sample C), as well as Nb-doped p-type samples A and B were found to exhibit good structural and material qualities as evident from their Raman spectra and high resolution XRD scans while transmission electron micrographs (TEM) showed good quality $MoS_2$ layers along (0001) orientation. Fig. 5 shows the on-axis XRD scan for the samples A, B and C. The (002) peak of $MoS_2$ was observed in all the three scans although the intensity for sample A was lower, indicating degradation of  material quality under such high doping levels ($2x10^{21}$ $cm^{-3}$). Sample B exhibited high intensity (002), (004) and (006) peaks, indicative of good crystalline quality of $MoS_2$. Fig. 6 shows the Raman spectra of of the samples taken with a Renishaw, 514 nm laser (with 60mW power). Characteristic in-plane ($E^1_{2g}$) and out-of-plane ($A_{1g}$) vibrational modes were observed at 381-382 and 407-408 $cm^{-1}$, respectively for all the three samples indicating crystalline nature of $MoS_2$. Atomic Force Micrograph (AFM) of sample B shown in Fig. 8 (rms roughness ~ 1.3 nm, scans size: 2 µm x 2 µm, data scale: 6 nm) indicated complete coverage of surface by $MoS_2$ and a relatively smooth surface. Fig. 7 shows cross sectional TEM image of undoped $MoS_2$ (control sample C) revealing ordered crystalline $MoS_2$, showing perfect stacking of $MoS_2$ layers along the (001) orientation.

Optical absorbance measurements were performed on all the samples using a broad UV-VIS-NIR deuterium-tungsten-halogen white light source. The absorbance spectra is taken for a reference piece of sapphire in order to determine a reference intensity $I_0(\lambda)$. The $MoS_2$ sample was measured and the absorbance was determined using $A = \frac{I(\lambda)}{I_0(\lambda)}$ where $I(\lambda)$ is the intensity collected by the monochromator after the light is transmitted through the $MoS_2$ grown on sapphire. The absorbance spectra (Fig. 7) shows that both undoped and Nb-doped $MoS_2$ (sample A) exhibit an absorption edge at 1.8 eV indicative of a direct band-to-band transition in these semiconductors. The absorption edge at 1.8 eV is zoomed in the inset to Fig. 7. The multiple blue/UV peaks are possibly due to transitions involving higher bands.

In conclusion, we show that Nb can act as an efficient acceptor in $MoS_2$ leading to high hole density and relatively high mobility. For a hole concentration of 4 x $10^{20}$ $cm^{-3}$, a hole mobility of 8.5 $cm^2/Vs$ was measured at room temperature, and was found to be limited by ionized impurity limited scattering. This use of Nb substitutional impurity for p-type doping demonstrated here for $MoS_2$ could be extended to other dichalcogenides, and could therefore have wider applications. Furthermore, the simple deposition scheme used here could be employed to directly pattern areas with p-type $MoS_2$, thus providing flexibility for device design. This first demonstration of substitutional p-type doping in large area thin film CVD-grown $MoS_2$ is expected to enable several high-performance electrical and opto-electronic devices.

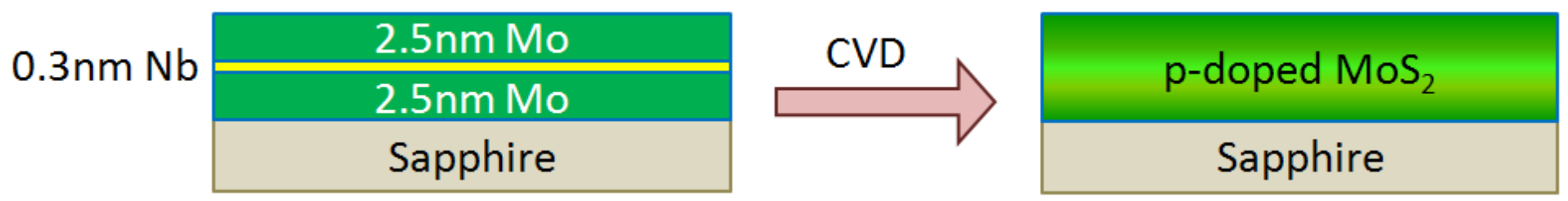

Fig. 1: Schematic of Nb-doping in CVD grown $MoS_2$

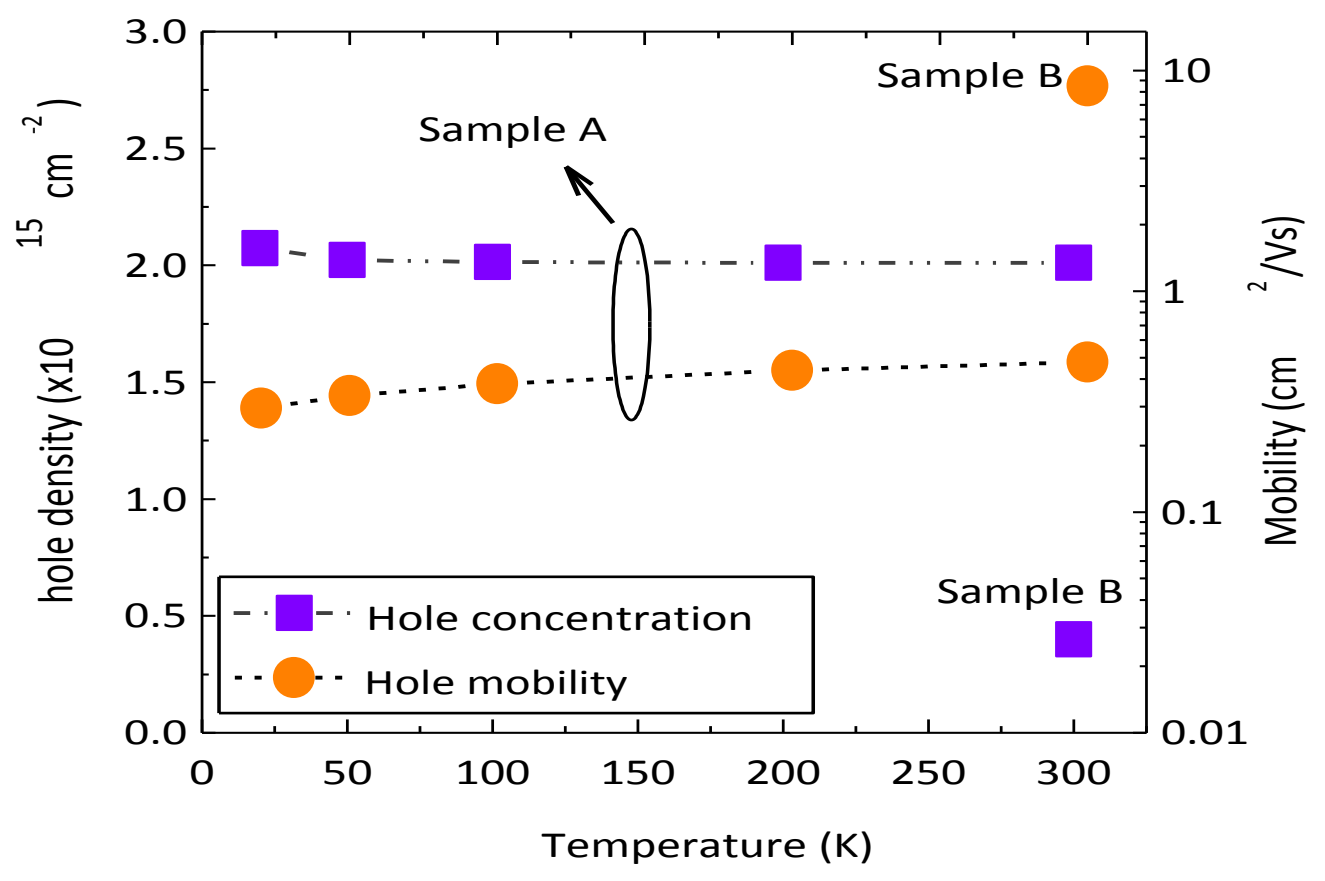

Fig. 2: Temperature-dependent hole mobility and charge density on sample A, compared with room temperature mobility and charge density for sample B.

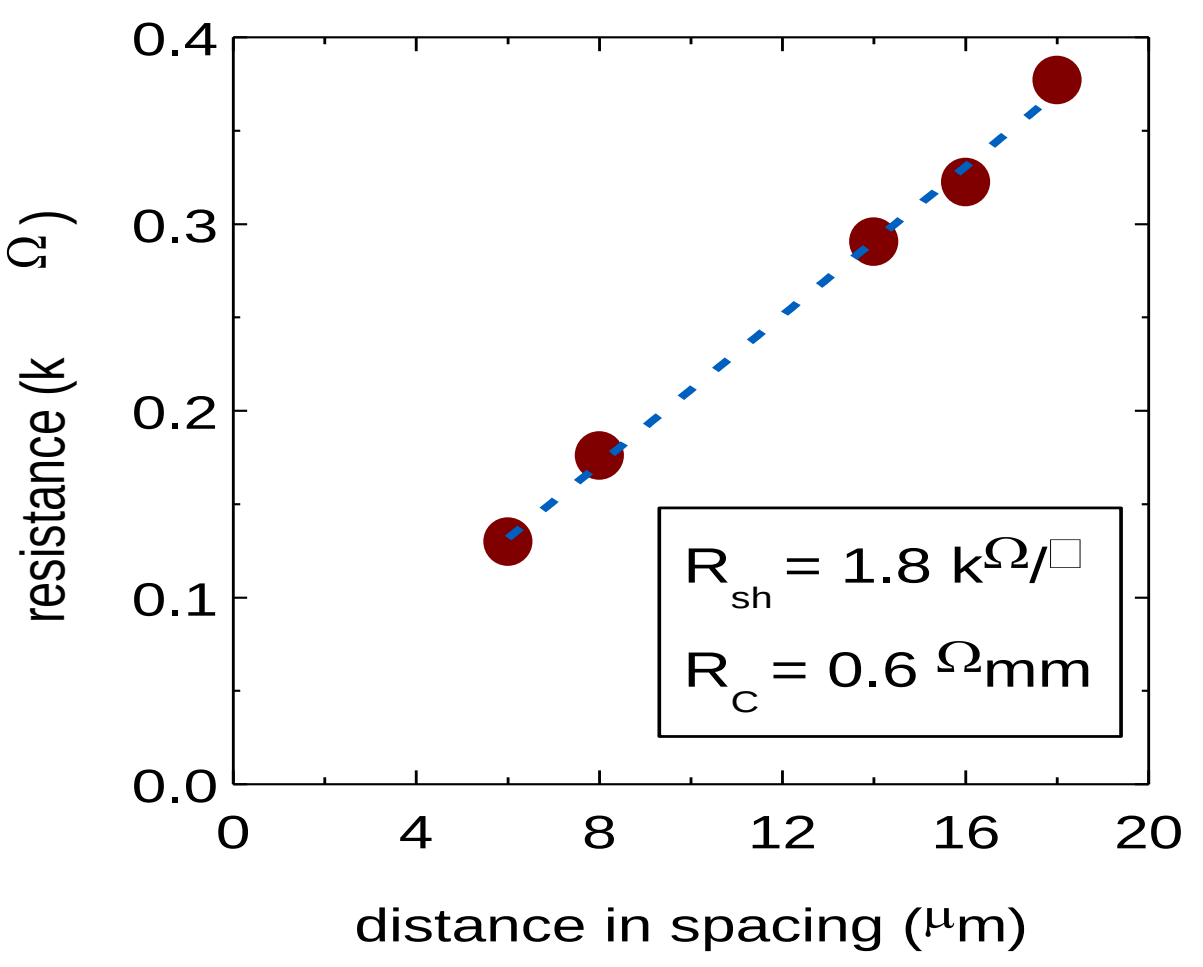

Fig. 3: TLM fitting to extract sheet and contact resistance on sample B using Ni/Au/Ni metal contacts

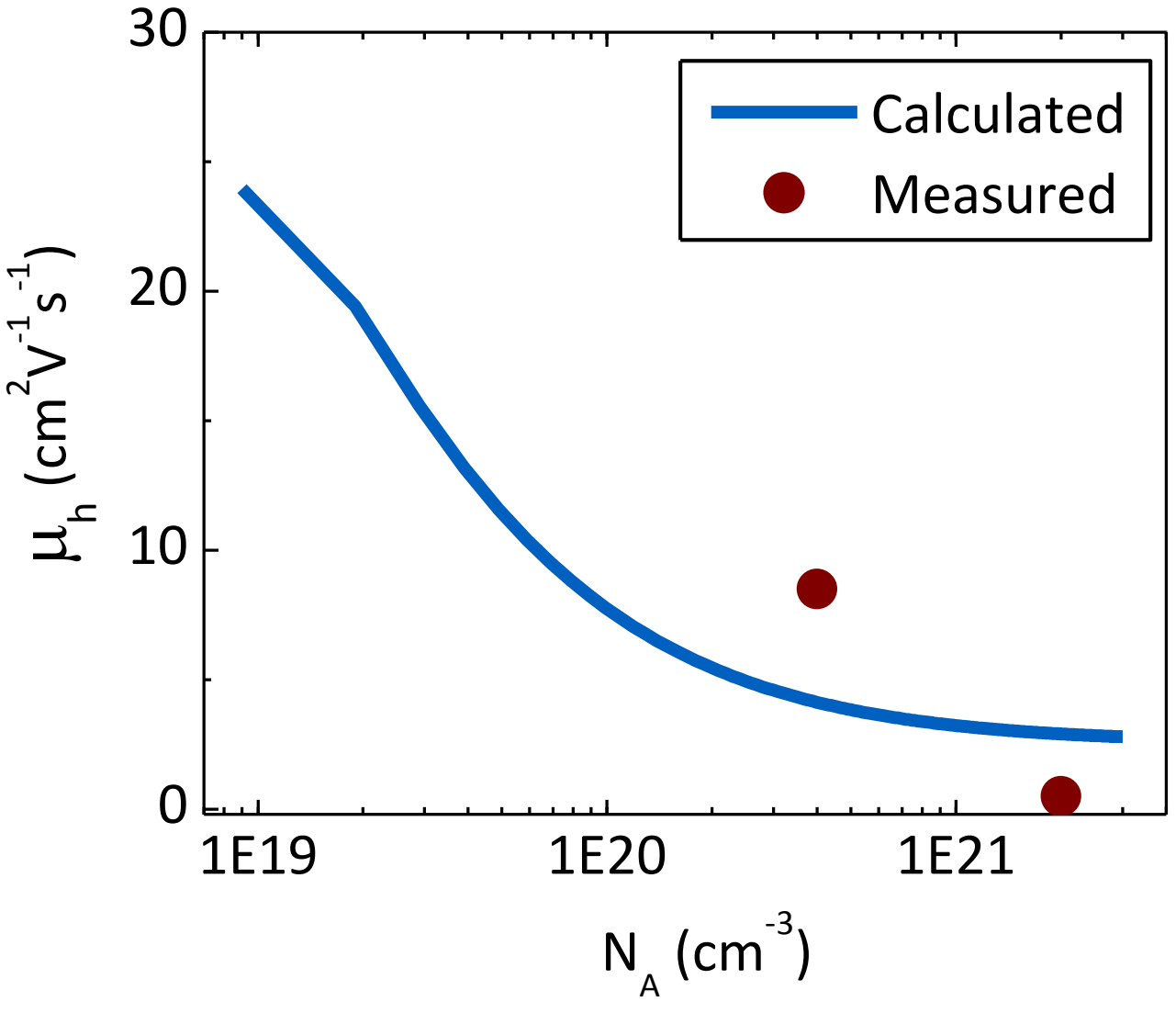

Fig. 4: Ionized impurity scattering limited hole mobility in p-$MoS_2$, theoretically estimated using 3D formalism, compared with measured data

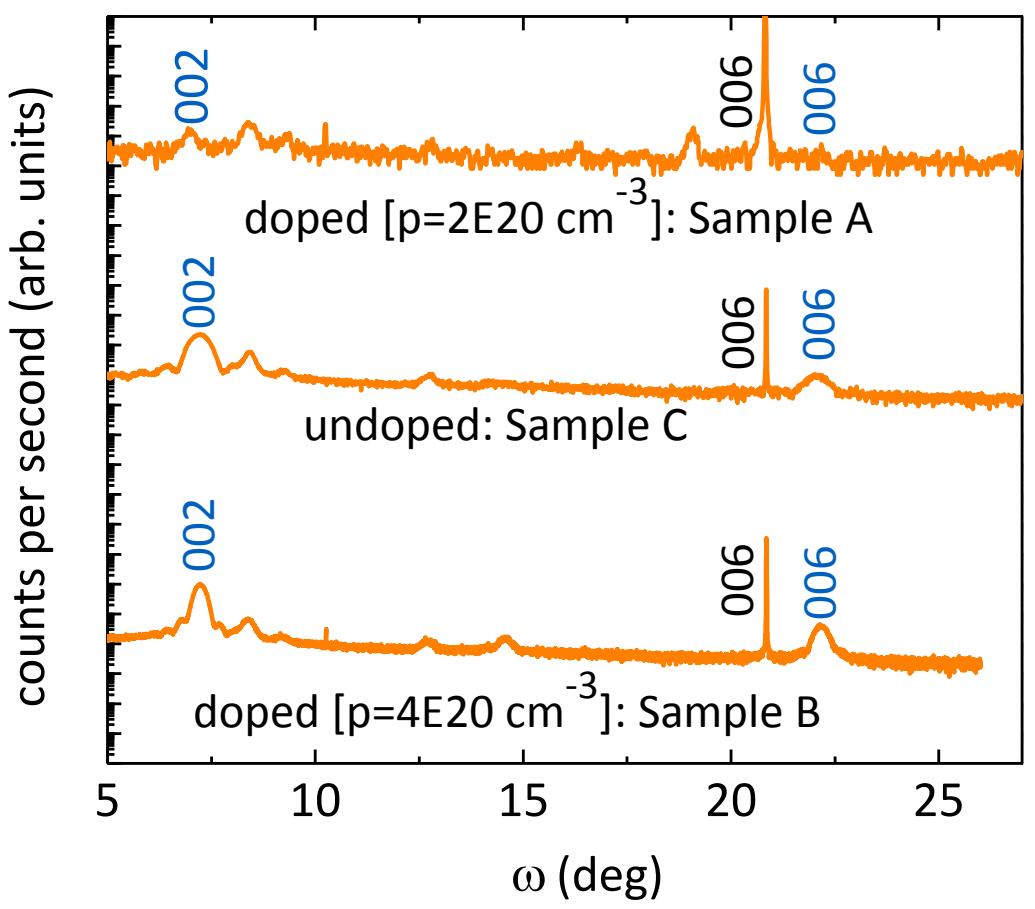

Fig. 5: High-resolution XRD scans for samples A, B and C

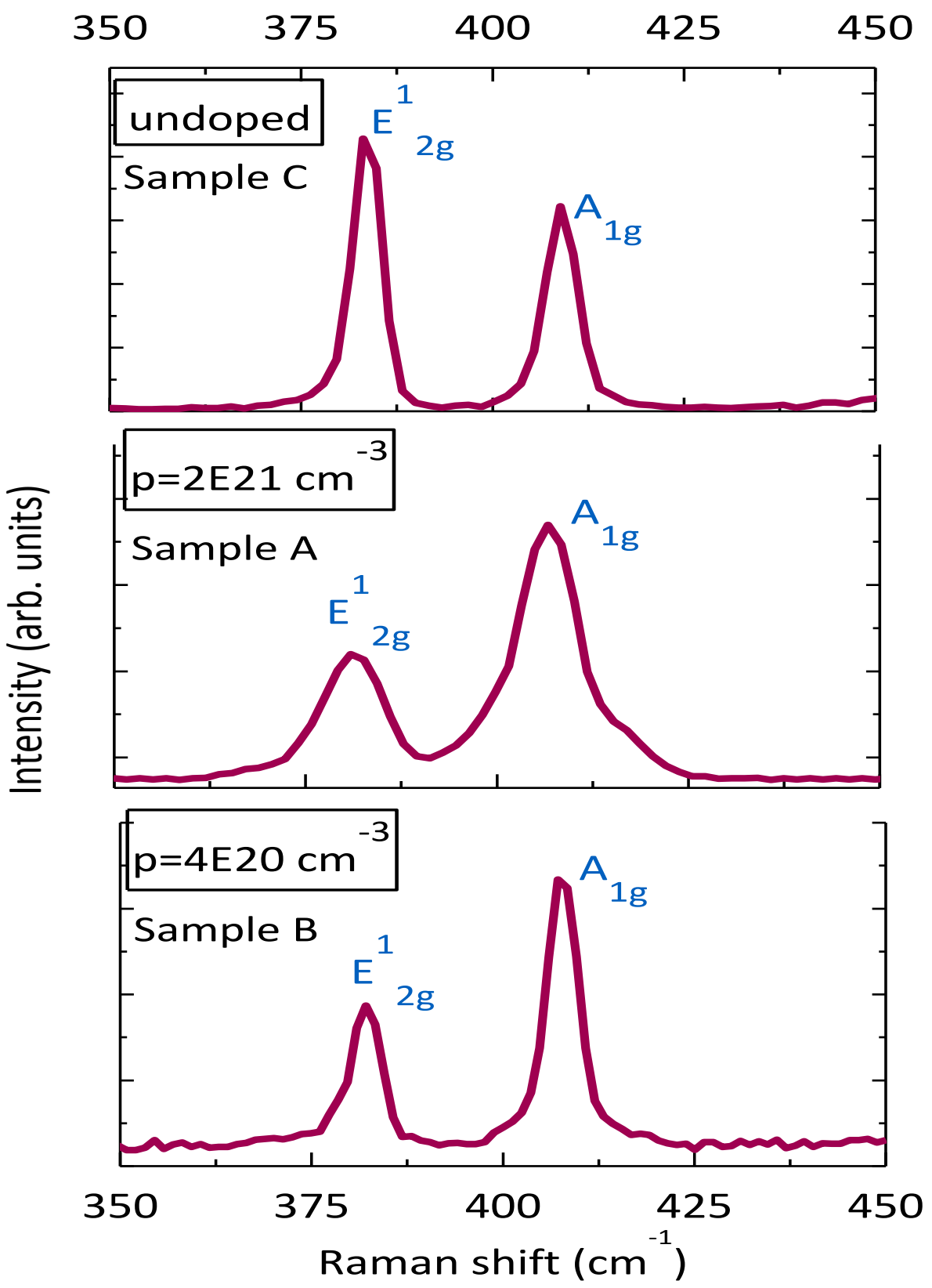

Fig. 6: Raman spectra on samples A, B, C

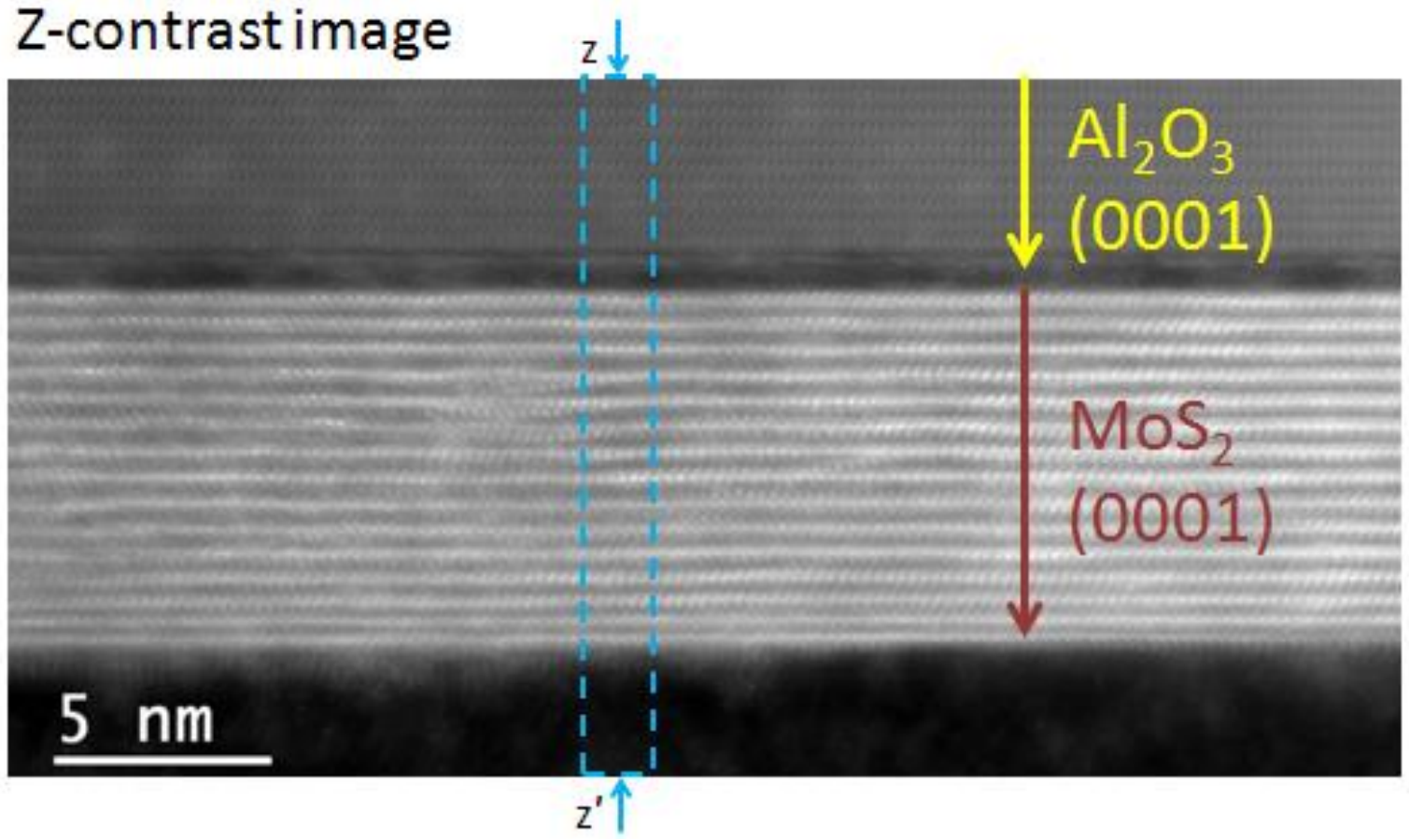

Fig. 7: Cross-sectional TEM image of undoped sample (C) showing perfect stacking of $MoS_2$ layers oriented along (0001) direcion.

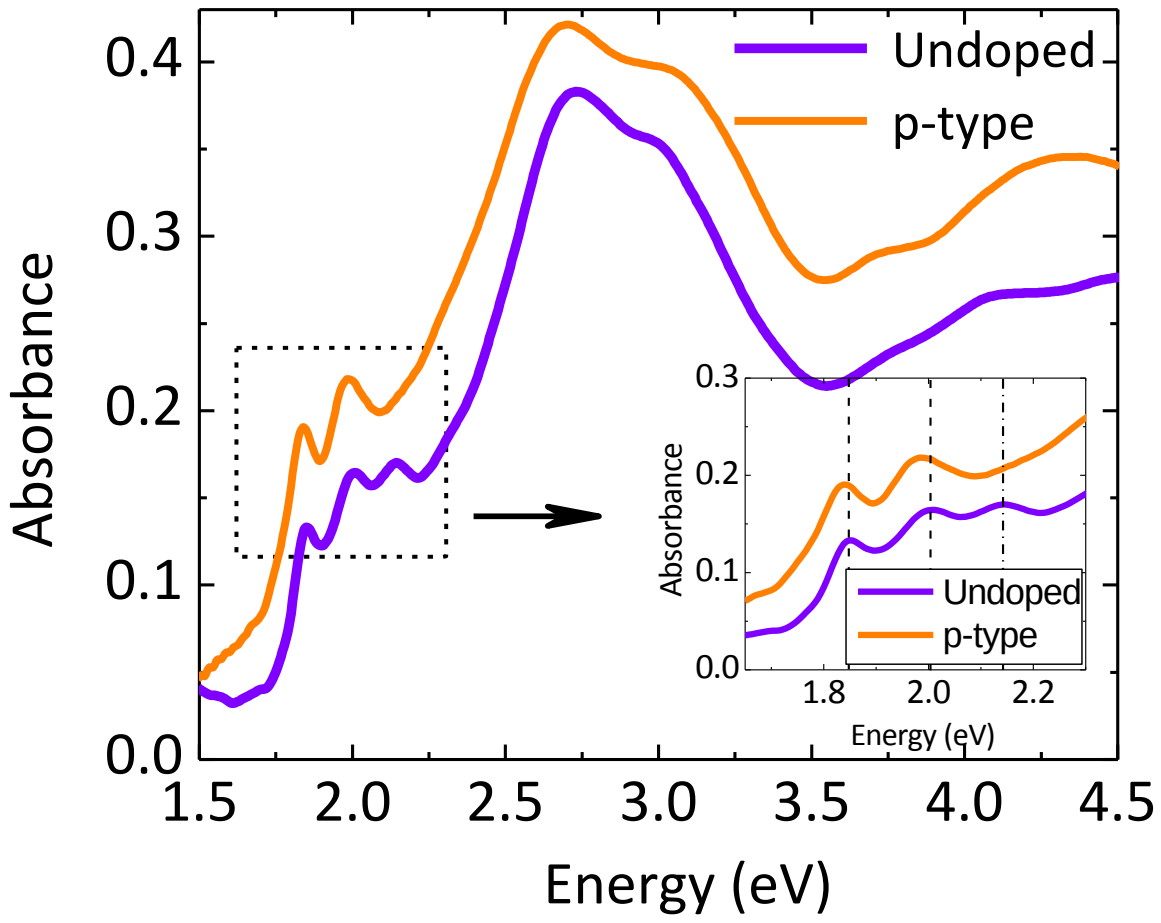

Fig. 8: Absorbance measurements for sample A and undoped sample C

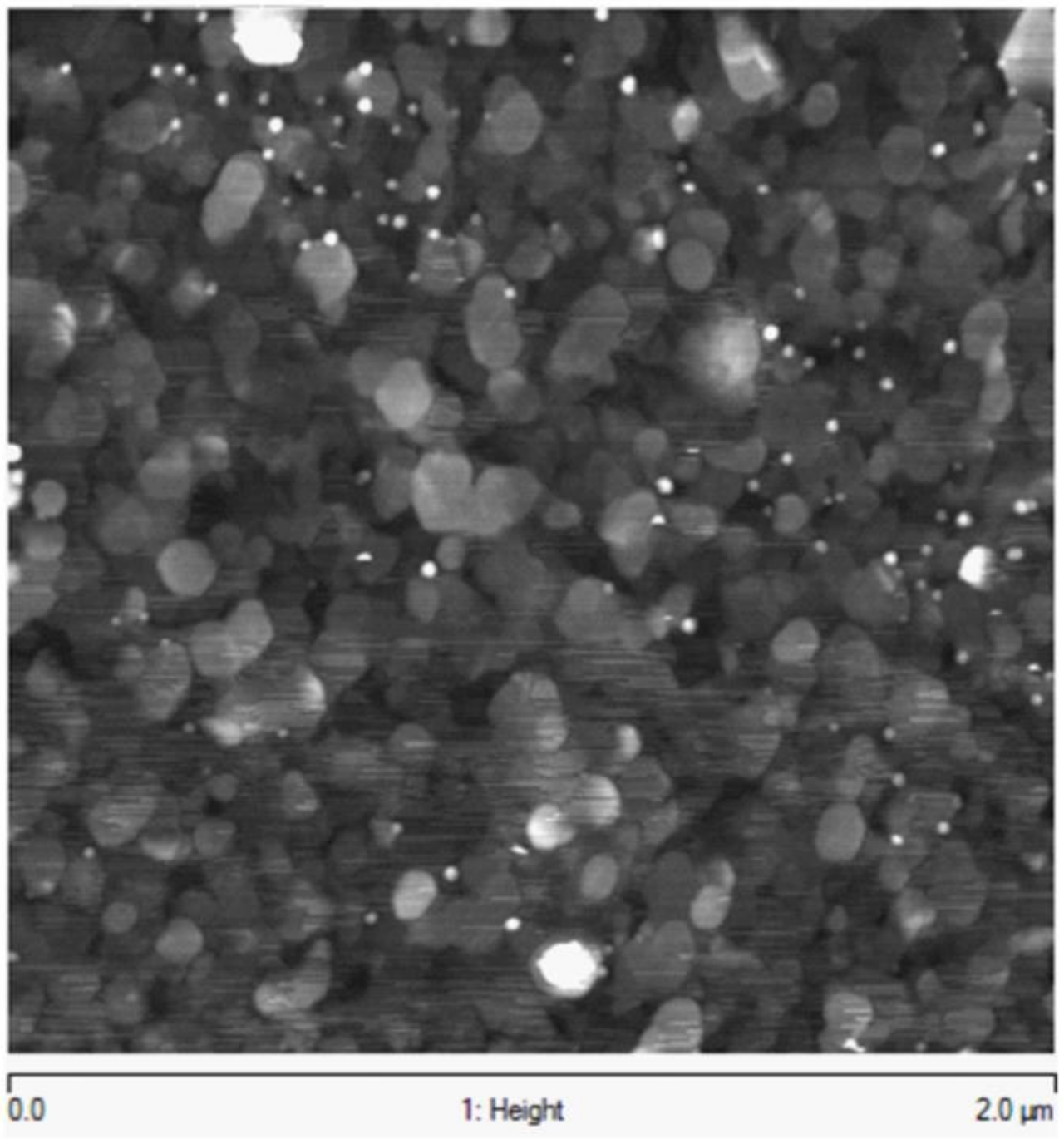

Fig. 9: AFM of sample B (2 µm x 2 µm scan), data scale: 8 nm; RMS roughness: 1.3 nm